\documentclass[conference]{IEEEtran}
\IEEEoverridecommandlockouts
\usepackage{cite}
\usepackage{amsmath,amssymb,amsfonts}
\usepackage{algorithmic}
\usepackage{graphicx}
\usepackage{textcomp}
\usepackage{amsmath}
\usepackage{booktabs} 
\usepackage{subcaption}
\usepackage[table]{xcolor}
\usepackage{mathtools}
\usepackage{tabularx}
\usepackage{array}
\usepackage{ragged2e}
\usepackage{makecell}
\usepackage{threeparttable}
\usepackage{dblfloatfix} 

\newcolumntype{P}[1]{>{\RaggedRight\arraybackslash}p{#1}}

\usepackage{tikz}
\usepackage{soul, color, xcolor}
\usepackage{hyperref}
\hypersetup{
    colorlinks,
    citecolor=blue,
    filecolor=blue,
    linkcolor=blue,
    urlcolor=blue
}
\usepackage{amsmath, nccmath}
\usetikzlibrary{shapes, arrows, positioning}
\def\BibTeX{{\rm B\kern-.05em{\sc i\kern-.025em b}\kern-.08em
    T\kern-.1667em\lower.7ex\hbox{E}\kern-.125emX}}
    \usepackage[a4paper, top=1.3cm, bottom=2.2cm, left=1.3cm, right=1.3cm]{geometry}
\begin{document}

\title{Design of Grid‑Forming Multi–Timescale Coordinated Control Strategies for Dynamic Virtual Power Plants \vspace{-0.3cm} \thanks{This study is supported by the State Key Laboratory of Alternate Electrical Power System with Renewable Energy Sources (Grant No.LAPS25006).}} 
\author{
\IEEEauthorblockN{Yan Tong\textsuperscript{1}, Qin Wang\textsuperscript{1}, Sihao Chen\textsuperscript{2}, Xue Hu\textsuperscript{1}, Zhaoyuan Wu\textsuperscript{3}}
\IEEEauthorblockA{
\textsuperscript{1}Department of Electrical and Electronic Engineering, The Hong Kong Polytechnic University, Hong Kong SAR\\
\textsuperscript{2}School of Electronic Information and Communications,
Huazhong University of Science and Technology, Wuhan, China\\
\textsuperscript{3}State Key Laboratory of Alternate Electrical Power System with Renewable Energy Sources,\\
North China Electric Power University, Beijing, China\\
qin-ee.wang@polyu.edu.hk
}\vspace{-0.6cm}}
\maketitle

\begin{abstract}
As the penetration level of distributed energy resources (DERs) continues to rise, traditional frequency and voltage support from synchronous machines declines. This weakens grid stability and increases the need for fast and adaptive control in a dynamic manner, especially in weak grids. However, most virtual power plants (VPPs) rely on static aggregation and plan-based resource allocation strategies. These methods overlook differences in device response times and limit flexibility for ancillary services. To address this issue, we propose a dynamic virtual power plant (DVPP) that coordinates heterogeneous resources across multiple time scales using grid-forming control. We first contrast grid-following and grid-forming converters: grid-following designs rely on a phase-locked loop which can undermine stability in weak grids, whereas our DVPP applies virtual synchronous generator control at the aggregate level to provide effective inertia and damping. Then, we introduce a dynamic participation factor framework that measures each device’s contribution through the frequency–active power and voltage–reactive power loops. Exploiting device heterogeneity, we adopt a banded allocation strategy: slow resources manage steady-state and low-frequency regulation; intermediate resources smooth transitions; and fast resources deliver rapid response and high-frequency damping. Comparative simulations demonstrate that this coordinated, timescale-aware approach enhances stability and ancillary service performance compared to conventional VPPs.
\end{abstract}


\section{Introduction}

\IEEEPARstart{A}{s} the penetration level of distributed energy resources (DERs) continues to rise, ancillary services traditionally ensured by conventional thermal units are increasingly being shouldered  by DERs. Because renewable DERs are variable and their available capacity changes over time, future power systems will require more robust frequency control and voltage regulation to maintain stability.

In recent years, virtual power plants (VPPs) that aggregate the DERs have gained increasing attention as a means to deliver ancillary services and grid control \cite{b1}. The VPP represents a new paradigm for energy management and dispatch. It leverages information and communication technologies, automatic control, and intelligent algorithms to logically integrate distributed, multi-type DERs into a unified and controllable equivalent resource. Although these assets are geographically dispersed, they are integrated via the VPP platform into a single “entity” at the logical level, which can participate in electricity markets, receive dispatch instructions, and provide energy and ancillary services much like a conventional large power plant \cite{b4}. The core value of a VPP lies in unlocking the “aggregation effect” of distributed resources, which significantly enhances system flexibility, economic efficiency, and reliability. Conventional VPPs rely on relatively static aggregation, with resource allocation based on preset plans or historical data. A dynamic VPP (DVPP) extends this model by adding real-time intelligence and adaptive coordination, allowing operating strategies to adjust continuously to grid conditions, market signals, weather, and load changes across distributed generation and storage \cite{b2}. A DVPP coordinates heterogeneous DERs within their physical and operational constraints to deliver fast, system-level frequency and voltage control \cite{b3}.

Existing DVPP control designs predominantly rely on aggregations of grid-following (GFL) DERs: frequency and voltage at the point of common coupling (PCC) are measured via a phase-locked loop (PLL) and used to modulate device outputs. Due to this reliance on measured frequency, GFL-based DVPPs generally require a stiff grid to operate stably; under high DER penetration or weak-grid conditions, their responsiveness and PLL tracking performance may degrade substantially and even trigger instability \cite{b5}.

Aligned with the growing use of grid-forming (GFM) resources, this paper proposes a DVPP architecture built on GFM units to deliver robust dynamic ancillary services. To make the DVPP function as a single power plant from the grid’s perspective, it must coordinate wind, solar, storage, and controllable loads—each with different response times—across both time and location. This aggregation is not a simple sum of outputs; it is a layered, timescale-aware, and constraint-driven coordination scheme. The response characteristics of major resource types are summarized in Table \ref{tab1}. At the aggregation layer, the DVPP should present its overall, externally visible dynamic capability to the market and system operator. Allocation and performance assessment are then carried out at this aggregate level, rather than through device-by-device commands.

\begin{table*}[!t]
    \renewcommand{\arraystretch}{1.25}
    \caption{Typical Response Characteristics of DVPP Assets}
    \label{tab:dvpp-dynamic-response}
    \centering
    \begin{threeparttable}
    \setlength{\tabcolsep}{2pt}
    \footnotesize
    \begin{tabular}{>{\centering\arraybackslash}p{3.0cm}|%
                    >{\raggedright\arraybackslash}p{4cm}%
                    >{\centering\arraybackslash}p{2.7cm}%
                    >{\centering\arraybackslash}p{3.3cm}%
                    >{\raggedright\arraybackslash}p{4cm}}
        \hline\hline
        \textbf{Asset Type} &
        \textbf{Dynamic Characteristics} &
        \textbf{Response Speed} &
        \textbf{Adjustable Time Horizon} &
        \textbf{Primary Control Objective} \\
        \hline
        Wind farm &
        Strongly correlated with wind speed, pronounced stochasticity &
        Minutes to hours &
        Mid-term (5 minutes to hours) &
        Power smoothing, production forecasting \\
        Photovoltaic plant &
        Correlated with solar irradiance, pronounced diurnal cycles &
        Seconds to minutes &
        Short to mid-term (seconds to hours) &
        Maximize energy yield, mitigate power fluctuations \\
        Energy storage system &
        Capable of rapid charge/discharge, highly agile &
        Milliseconds to seconds &
        Very short-term (milliseconds to minutes) &
        Frequency regulation, peak shaving, emergency reserve \\
        Controllable load / demand response &
        Dependent on end-user behavior, inherent delays &
        Minutes to hours &
        Medium to long term &
        Peak-valley balancing, load shifting \\
        Microgrid / virtual power plant &
        Coordinated control enabled, flexible operational strategies &
        Seconds to minutes &
        Short to mid-term &
        Power coordination, backup provision \\
        \hline\hline
    \end{tabular}
    \begin{tablenotes}[flushleft]
        \footnotesize
        \item \textit{Note:} Renewable generation (wind and solar) evolves comparatively slowly, storage responds the fastest, whereas loads and microgrids bridge the intermediate temporal layers.
    \end{tablenotes}
    \end{threeparttable}
        \label{tab1}
\end{table*}

The main contributions of this paper are summarized as follows: 1) We propose a DVPP framework that combines GFM capability with dynamic aggregation. By introducing virtual synchronous generator (VSG) control at the aggregation layer, the DVPP provides quantifiable effective inertia and damping, thereby improving stability margins under weak-grid and high-penetration conditions. 2) We develop a coordination strategy centred on dynamic participation factors that explicitly accounts for time-varying resource capacities and disparate response timescales. The strategy enables bandwidth-aware task allocation from milliseconds to hours and aligns the aggregated external behaviour with system-level requirements, offering a deployable control paradigm for ancillary services. 3) To facilitate reproducibility, we provide the algorithms required for the experimental design as well as the datasets used in the experiments.

In summary, this paper develops a DVPP architecture supported by GFM control. Section II presents the control design and implementation pathway for GFM inverters within the DVPP. Section III proposes a dynamic participation factor framework for the coupled \(\omega -P\) power and \(v-Q\) power channels. Guided by device-level physical constraints and timescale characteristics, the desired aggregate behavior is decomposed across wind/PV, energy storage, and controllable loads, and multi–timescale coordination is achieved via low-/high-/band-pass characteristics. Section IV conducts three comparative studies on a modified 3-machine, 9-bus system to validate the effectiveness of the proposed method in terms of frequency coherency, Rate Of Change Of Frequency (ROCOF) suppression, enhancement of effective inertia and damping, and feasible decoupling and limit handling for active/reactive power.

\section{GFM Control Design for DVPP}
\subsection{GFL and GFM}
\begin{figure}[htbp]
    \centering
    \begin{subfigure}[b]{0.45\textwidth}
        \centering
        \includegraphics[width=\linewidth]{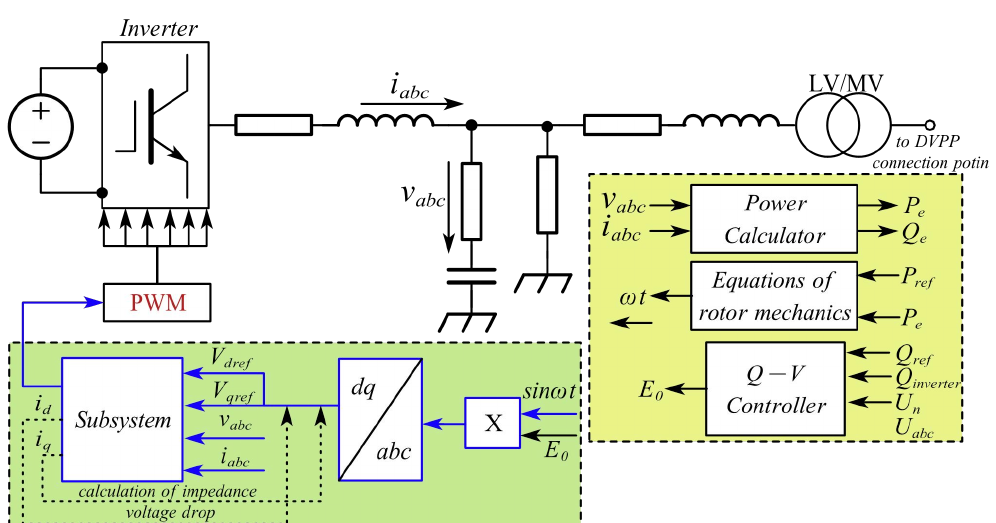}
        \caption{}
        \label{1(A)}
    \end{subfigure}
    \hfill
    \begin{subfigure}[b]{0.45\textwidth}
        \centering
        \includegraphics[width=\linewidth]{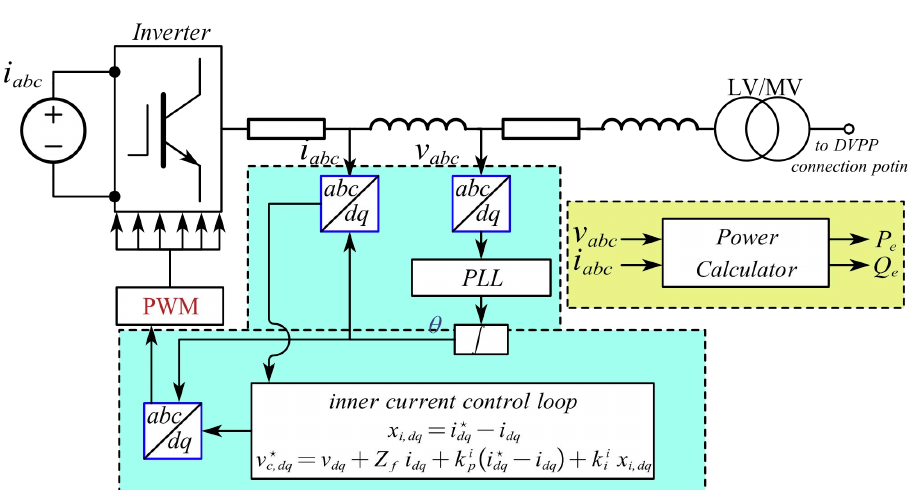}
        \caption{}
        \label{1(B)}
    \end{subfigure}
    \caption{Inverters and their control structures. (a) Grid-forming inverter. (b) Grid-following inverter.}
    \label{1}
\end{figure}

Power electronic inverters for DERs are commonly divided into two categories \cite{b6}, as illustrated in Fig.\ref{1}: GFL and GFM. 
A GFL inverter employs a PLL to track the phase angle of the existing grid voltage and subsequently regulates the AC-side current. 
In general, a GFL inverter consists of three control loops, namely the outer-loop power controller, the inner-loop current controller, and the PLL:

\begin{itemize}
    \item Outer-loop power controller:
    \begin{equation}
        \begin{aligned}
            P = \dfrac{3}{2} \left( v_{gd} i_{od} + v_{gq} i_{oq} \right),
            \quad
            Q = \dfrac{3}{2} \left( v_{gq} i_{od} - v_{gd} i_{oq} \right)
        \end{aligned}
        \label{q1}
    \end{equation}

    \item Inner-loop current controller:
    \begin{small} 
    \begin{equation}
        \begin{aligned}
            v_{id}^{ref} &= v_{gd} + L_f \!\left[k_p (i_d^{ref} - i_d) + k_i 
            \!\int (i_d^{ref} - i_d)\, dt - \omega L_f i_q \right]\\[4pt]
            v_{iq}^{ref} &= v_{gq} + L_f \!\left[k_p (i_q^{ref} - i_q) + k_i 
            \!\int (i_q^{ref} - i_q)\, dt + \omega L_f i_d \right]
        \end{aligned}
        \label{q2}
    \end{equation}
    \end{small}

    \item Phase-Locked Loop (PLL):
    \begin{equation}
        \dot{\hat{\theta}} = \omega_{pll} = K_p e_v + K_i \int e_v \, dt
        \label{q3}
    \end{equation}
\end{itemize}
where $P$ and $Q$ are the instantaneous active and reactive powers; $v_{gd}$ and $v_{gq}$ are the $d$- and $q$-axis grid voltages in the synchronous frame; $i_{od}$ and $i_{oq}$ are the inverter output currents on the $d$–$q$ axes; $v_{id}^{\mathrm{ref}}$ and $v_{iq}^{\mathrm{ref}}$ are the reference inverter voltages on the $d$–$q$ axes; $i_{d}^{\mathrm{ref}}$ and $i_{q}^{\mathrm{ref}}$ are the $d$–$q$ reference currents; $i_d$ and $i_q$ are the measured currents; $L_f$ is the inverter-side filter inductance; $k_p$ and $k_i$ are the PI current-controller gains; $\omega$ is the grid angular frequency; $\hat{\theta}$ is the grid-voltage phase estimated by the PLL; $\omega_{\mathrm{pll}}$ is the PLL-estimated angular frequency; $K_p$ and $K_i$ are the PLL gains; and $e_v$ is the PLL phase-detection error.

GFM inverters have been widely deployed to integrate wind and solar power into the grid owing to their simple control architecture, the maturity of PLL technology, and their operation under prescribed current \cite{b7}. However, the PLL can adversely affect system stability, particularly when the grid is weak due to high grid impedance. As  Synchronous Generators (SGs) are increasingly replaced by Inverter-Based Resources (IBRs), this issue is likely to arise more frequently and become more challenging worldwide. Consequently, attention has shifted in recent years toward GFM inverters. GFM inverters regulate the AC-side voltage and contribute to establishing a voltage-source-type grid. They synchronise with the rest of the grid via frequency Droop (DP) control, in a manner analogous to SGs.

\subsection{Virtual Synchronous Generator}

DP control is currently the most common GFM strategy, as it links the output power to the system frequency, thereby emulating the speed-DP characteristic of SGs \cite{b8}. However, with system stability in mind, implementing GFM voltage–power control using only \ref{q2} and \ref{q3} is insufficient. 

\begin{equation}
\omega  = {\omega _{ref}} - K_P^{droop}(P - {P_{ref}}),
\quad
v = {v_{ref}} - K_d^{droop}(Q - {Q_{ref}})\
\label{q4}
\end{equation}
where \(\omega\) represents the measured angular frequency of the inverter's output voltage, \(v\) denotes the measured output voltage of the inverter, \(P\) is the active power output of the inverter, \(Q\) is the reactive power output, \(P_{ref}\) is the rated reference active power, \(Q_{ref}\) is the rated reference reactive power, \(\omega _{ref}\) is the rated reference angular frequency, \(v_{ref}\) is the reference voltage value,  \(K_P^{droop}\) is the DP coefficient for active power, and \(K_d^{droop}\) is the DP coefficient for reactive power.

In \cite{b9}, a low-pass filter is introduced to attenuate high-frequency harmonic disturbances in the measurements. Its transfer function is given by:

\begin{equation}
\omega  = {\omega _{ref}} - \frac{1}{{\tau s + 1}}K_P^{droop}(P - {P_{ref}})\label{q5}
\end{equation}
\begin{equation}
v = {v_{ref}} - \frac{1}{{\tau s + 1}}K_d^{droop}(Q - {Q_{ref}})\label{q6}
\end{equation}
where \(\tau\) represents the time constant of the low-pass filter, which is typically defined as \(\tau  = 1/{\omega _c}\), where \(\omega _c\) is the cutoff angular frequency of the low-pass filter.

The VSG strategy is an increasingly popular inverter control approach \cite{b10}. Most VSG strategies emulate the swing equation of SG by leveraging short-term energy storage and appropriate control algorithms.\cite{b9} proposes a generalized GFM control scheme that endows the inverter with both DP and VSG characteristics. Across these methods, the rotor inertia is typically the same, whereas the forms of the damping characteristics differ. The \(\omega -P\) DP controller (equation \ref{q5} and the mechanical equation of a sg (equation \ref{q6} can be expressed, respectively, as:
\begin{equation}
{P_m} = {P_{ref}} + \frac{1}{{\tau s + 1}}{K_\omega }({\omega _{ref}} - \omega ),
\quad
\ddot \theta  = \frac{1}{J}({T_m} - {T_e} - {T_d})
\label{q7}
\end{equation}
where \(T_m\) represents the mechanical torque, \(T_e\) denotes the electromagnetic torque, \(T_d\) is the damping torque, \(J\) refers to the moment of inertia, \(D\) is the damping coefficient, and \(P_m\) represents the mechanical power.
When the damping torque is expressed as \(D \cdot (\omega  - {\omega _0})\) \cite{b11}, the derivation has been presented in prior work \cite{b9} and is therefore omitted here. The resulting system transfer function is given by:
\begin{equation}
\omega  = \frac{1}{J} \cdot \frac{1}{s} \cdot \frac{{{P_{ref}} + \frac{1}{{\tau s + 1}}{K_\omega }({\omega _{ref}} - \omega ) - P}}{\omega } - \frac{{D(\omega  - {\omega _0})}}{{Js}} + {\omega _0}\label{q8}
\end{equation}

When devices in a DVPP are connected in parallel to the grid as GFMs employing VSG control, and coordinated via appropriate control strategies, PV, energy storage, and wind units emulate SG behavior. They introduce virtual inertia \(J\) and damping \(D\) to the system, thereby smoothing frequency variations, providing transient energy buffering, and more effectively suppressing power and frequency oscillations.

\section{Design of DVPP Output Coordination Methods}
\subsection{Definition of the DVPP Dynamic Participation Factor}
In the design of DPP, we consider a VPP as an aggregation of renewable energy devices within a given region. These devices include, but are not limited to, wind power, PV generation, BESS, and supercapacitors. Following the classification approach in \cite{b2}, we categorize all generation units in the model into two groups: controllable units based on GFM inverters, denoted by set \(X\); and uncontrollable units based on SGs—such as conventional thermal units, hydropower units, and certain wind power units—denoted by set \(Y\). When the system load power changes:

\begin{equation}
\begingroup
\setlength{\arraycolsep}{2pt}
\begin{aligned}
\Delta p_{1} + \Delta p_{2} + \Delta p_{3} + \cdots + \Delta p_{x}
&= \sum_{\mathclap{i \in X}} \Delta p_{i}\\[2pt]
\Delta q_{1} + \Delta q_{2} + \Delta q_{3} + \cdots + \Delta q_{x}
&= \sum_{\mathclap{i \in X}} \Delta q_{i}\\[2pt]
\Delta p'_{1} + \Delta p'_{2} + \Delta p'_{3} + \cdots + \Delta p'_{y}
&= \sum_{\mathclap{i \in Y}} \Delta p'_{i}\\[2pt]
\Delta q'_{1} + \Delta q'_{2} + \Delta q'_{3} + \cdots + \Delta q'_{y}
&= \sum_{\mathclap{i \in Y}} \Delta q'_{i}
\end{aligned}
\endgroup
\label{q9}
\end{equation}

\begin{equation}
\begingroup
\setlength{\arraycolsep}{2pt}
\begin{alignedat}{6}
\Delta p_{\text{target}} &= \sum_{\mathclap{i \in X}}\Delta p_i & + & \sum_{\mathclap{i \in Y}}\Delta p'_i
&\quad&
\Delta q_{\text{target}} &= \sum_{\mathclap{i \in X}}\Delta q_i & + & \sum_{\mathclap{i \in Y}}\Delta q'_i
\end{alignedat}
\endgroup
\label{q10}
\end{equation}

We assume that all uncontrollable devices are equipped with governors or equivalent regulation mechanisms. Furthermore, for controllable devices that are interfaced through GFM inverters, classical DP control implies a well-defined relationship between frequency and active power, as well as between voltage and reactive power. Introducing a participation factor \(\xi\), we then obtain the following relations for both controllable and uncontrollable devices:

\begin{equation}
\begingroup
\setlength{\arraycolsep}{2pt}
\begin{alignedat}{6}
\varDelta p_{target}\left( s \right) =\sum{_{i\in x\cup y}}\xi _i\left( s \right) \varDelta \omega \left( s \right) \\
\varDelta q_{target}\left( s \right) =\sum{_{i\in x\cup y}}\xi _i\left( s \right) \varDelta v\left( s \right) 
\end{alignedat}
\endgroup
\label{q11}
\end{equation}

For conventional, uncontrollable units such as thermal generators, the participation factor a is fixed. By contrast, for controllable units, a is subject to the following aggregation conditions:
\begin{equation}
\xi _{target}\left( s \right) =\sum{_{i\in x\cup y}}\xi _i\left( s \right) 
\label{q12}
\end{equation}

During the control design stage, we aim to account for the devices’ physical and operating constraints, including bandwidth, sensitivity to time-varying power fluctuations, and the requirement that currents remain within limits under normal operating conditions. Moreover, to enable aggregation across disparate timescales, each area of the system must contain a sufficient number of devices. Consequently, when the power system operator specifies the desired \(\xi _{target}\), we allocate to each device an appropriate participation factor a based on its capacity, timescale characteristics, and physical constraints. In this way, while ensuring coverage of the entire range of timescales, the aggregated DVPP can provide adequate system support and maintain overall stability. The notation used in the formulas of this section is provided in Table \ref{tab2}.

\begin{table}[htbp]
    \centering
    \scriptsize
    \setlength{\tabcolsep}{6pt}
    \renewcommand{\arraystretch}{1.35}
    \rowcolors{2}{gray!15}{white}
    \caption{Notation for DVPP Output Coordination Methods}
    \label{tab2}
    \begin{tabular}{>{\centering\arraybackslash}m{0.64\linewidth} >{\centering\arraybackslash}m{0.20\linewidth}}
        \hline
        \textbf{Description} & \textbf{Symbol} \\
        \hline\hline
        Coordinated grid-forming asset cluster within the DVPP & $X$ \\
        Synchronous-machine cluster treated as non-dispatchable & $Y$ \\
        Index labeling an individual DVPP constituent & $i$ \\
        \hline
        Active-power deviation supplied by controllable unit $i$ & $\Delta P_{i}$ \\
        Reactive-power deviation supplied by controllable unit $i$ & $\Delta Q_{i}$ \\
        Active-power deviation delivered by governor-driven unit $i$ & $\Delta P'_{i}$ \\
        Reactive-power deviation delivered by governor-driven unit $i$ & $\Delta Q'_{i}$ \\
        DVPP-level active-power support request from the operator & $\Delta P_{\mathrm{target}}$ \\
        DVPP-level reactive-power support request from the operator & $\Delta Q_{\mathrm{target}}$ \\
        Bus-frequency perturbation observed in the Laplace domain & $\Delta \omega(s)$ \\
        Bus-voltage-magnitude perturbation observed in the Laplace domain & $\Delta v(s)$ \\
        \hline
        Frequency-support participation weighting for unit $i$ & $\xi_{i}(s)$ \\
        Aggregate participation profile mandated by the operator & $\xi_{\mathrm{target}}(s)$ \\
        \hline
    \end{tabular}
\end{table}

\subsection{Design of the DVPP Dynamic Participation Factor}
To enable effective DVPP control design, we introduce dynamic participation factors(DPF) to decompose the system operator’s overall desired behavior into device-level targets, while explicitly accounting for each device’s characteristics. In this way, the resulting allocation effectively covers tasks across different timescales.

By applying the DPF, we distribute the overall desired behavior to each individual device as follows:
\begin{equation}
\varGamma \left( s \right) \cdot \xi _{target}\left( s \right) =\xi _i\left( s \right) , \forall i\in X\cup Y
\label{q13}
\end{equation}

Here, a denotes a matrix that specifies the allocation of DPF; its definition is given as follows:
\begin{equation}
\varGamma \left( s \right) =\left[ \begin{matrix}
	\varphi _{i}^{\omega p}&		0\\
	0&		\gamma _{i}^{vq}\\
\end{matrix} \right] , \forall i\in =X\cup Y
\label{q14}
\end{equation}
where \(\varphi _{I}^{\omega p}\) and \(\gamma _{i}^{vq}\) represent the DPF associated with the \(\omega -p\) and \(v-q\) channels, respectively.

Accordingly, we may reformulate equation \ref{q12} as follows:
\begin{equation}
\xi _{target}\left( s \right) =\sum{_{i\in x\cup y}}\varGamma \left( s \right) \cdot \xi _{target}\left( s \right) =\sum{_{i\in x\cup y}}\xi _i\left( s \right) 
\label{q15}
\end{equation}

It should be emphasized that the foregoing formulation holds only under the following prerequisite:
\begin{equation}
\sum{_{i\in x\cup y}}\varphi _{I}^{\omega p}\left( s \right) =1 ,
\quad
\,\,\sum{_{i\in x\cup y}}\gamma _{i}^{vq}\left( s \right) =1
\label{q16}
\end{equation}

For uncontrollable units such as synchronous generators already equipped with governors and related regulators, their DPFs are fixed and can be expressed as:

\begin{equation}
\varGamma \left( s \right) =\left( \xi _{target}\left( s \right) \right) ^{-1}\cdot \xi _i\left( s \right) , \forall i\in y
\label{q17}
\end{equation}

Given the pronounced heterogeneity among devices in power capacity, response time, and dynamic characteristics, a naive static proportional allocation or DPF would inevitably degrade performance at certain timescales and/or lead to underutilization of resources. Therefore, we propose introducing auxiliary control modules within the control loop to regulate the DPFs across devices.

In this context, we propose assigning device-type-specific participation functions. For units capable of sustaining power output over longer time horizons and providing steady-state regulation, the participation factor is designed as a low-pass filter. This choice ensures strong responsiveness at low frequencies while suppressing high-frequency participation, thereby preventing operation beyond their dynamic capabilities.

Conversely, for devices characterised by very fast response yet limited energy capacity, the participation factor is designed as a high-pass filter. Such devices primarily address high-frequency components, delivering rapid dynamic regulation to attenuate short-term disturbances.

For devices whose characteristics lie between these two extremes, a band-pass filter is adopted. This design fills the regulation gap at intermediate timescales and ensures a smooth transition of the aggregate system response across frequency bands.

\section{Experimental Simulation and Parameter Design}
\vspace{-0.2cm}
\begin{figure}[htbp]
\centering
\includegraphics[width=0.45\textwidth]{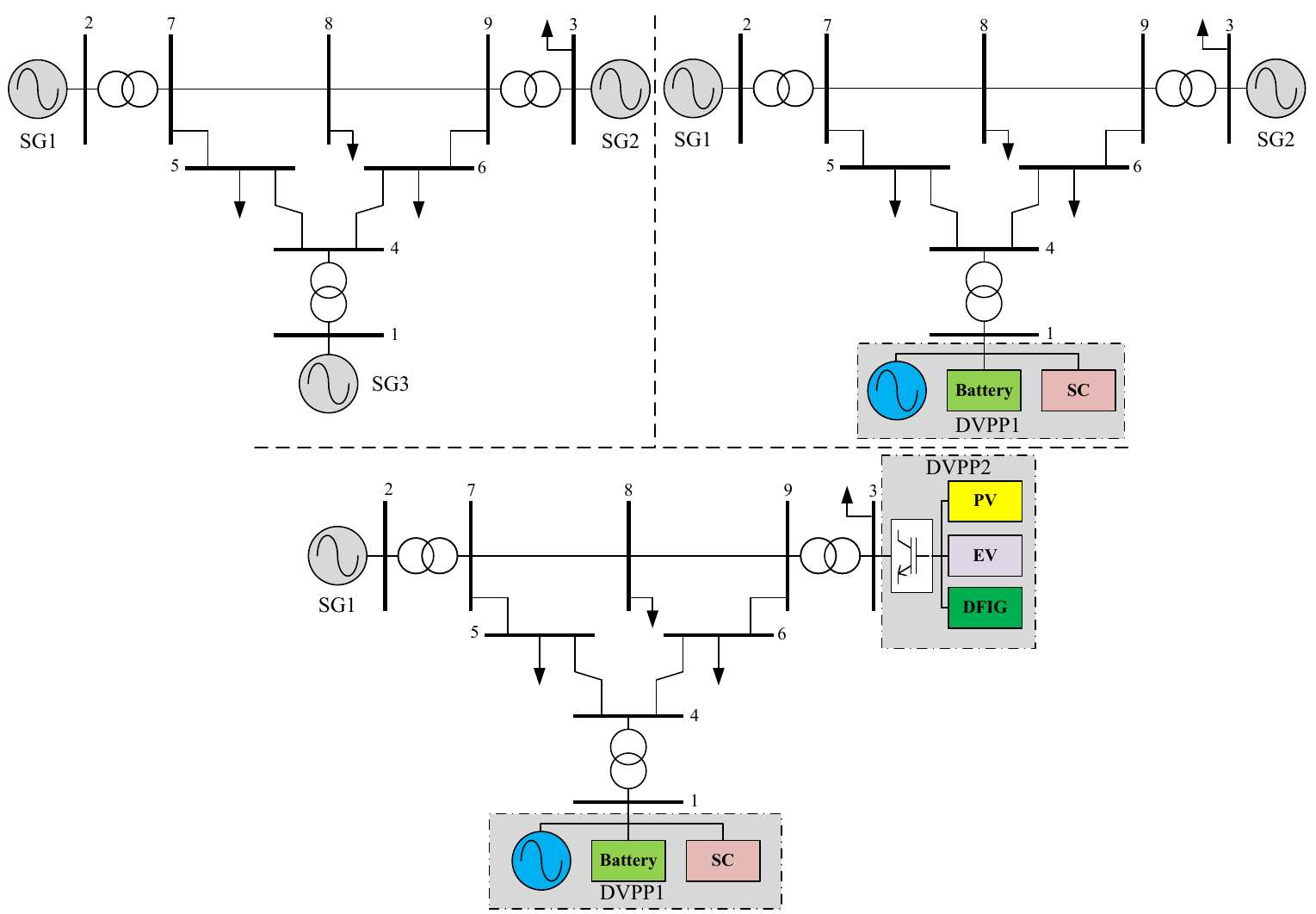}
\caption{IEEE nine-bus system}
\label{2}
\end{figure}

To validate the feasibility of the proposed DVPP control method, we conducted three experiments in PSCAD using the conventional 3-machine 9-bus IEEE system as the baseline, as shown in Fig.~\ref{2}. In principle, the method scales to larger distribution networks; moreover, because the adaptive dynamic participation mechanism is independent of the number of homogeneous devices within a DVPP, we model only one representative unit per type. In all three experiments, step changes in load during operation emulate demand fluctuations, allowing us to examine system frequency and power responses. Experiment I serves as the baseline with all three units modeled as SGs. In Experiment II, the third SG is replaced by DVPP1 comprising a large hydropower unit and a supercapacitor (SC), and the proposed adaptive control is applied to realize a banded coordination strategy—namely, fast support by the SC, mid-band regulation by the storage unit, and steady-state support by hydropower. Building on Experiment II, Experiment III replaces the second SG with DVPP2 consisting of PV, a DFIG-based wind unit, and vehicle-based storage, and adopts a VSG-based GFM control to enhance equivalent inertia and damping while maintaining multi–timescale coordination. Key parameters are summarized in Table \ref{tab3}.

\begin{table}[htbp]
\captionsetup{justification=centering, singlelinecheck=false}
\caption{Characteristic parameters of the system}
\centering
\renewcommand\arraystretch{1.4}
\setlength{\tabcolsep}{1.8mm}{
\begin{tabular}{cc|cc}
\hline
\hline
Parameter & Value & Parameter & Value \\
\hline
System base power $S$ & 100 MV & Power rating, DFIG & 70 MVA \\
System base voltage $v$ & 230 kv &  Power rating, BA & 50 MVA \\
Frequency $f$ & 60 HZ & Power rating, sc & 20 MVA \\
Power rating, SG1 & 250 MVA & Power rating, DVPP1 & 250 MVA \\
Power rating, SG2 & 80 MVA & Power rating, DVPP2 & 60 MVA \\
Power rating, SG3 & 60 MVA & Power rating, PV & 52 MVA \\
\hline
\hline
\end{tabular}}
\label{tab3}
\end{table}

\begin{figure}[htbp]
\centering
\includegraphics[width=0.48\textwidth]{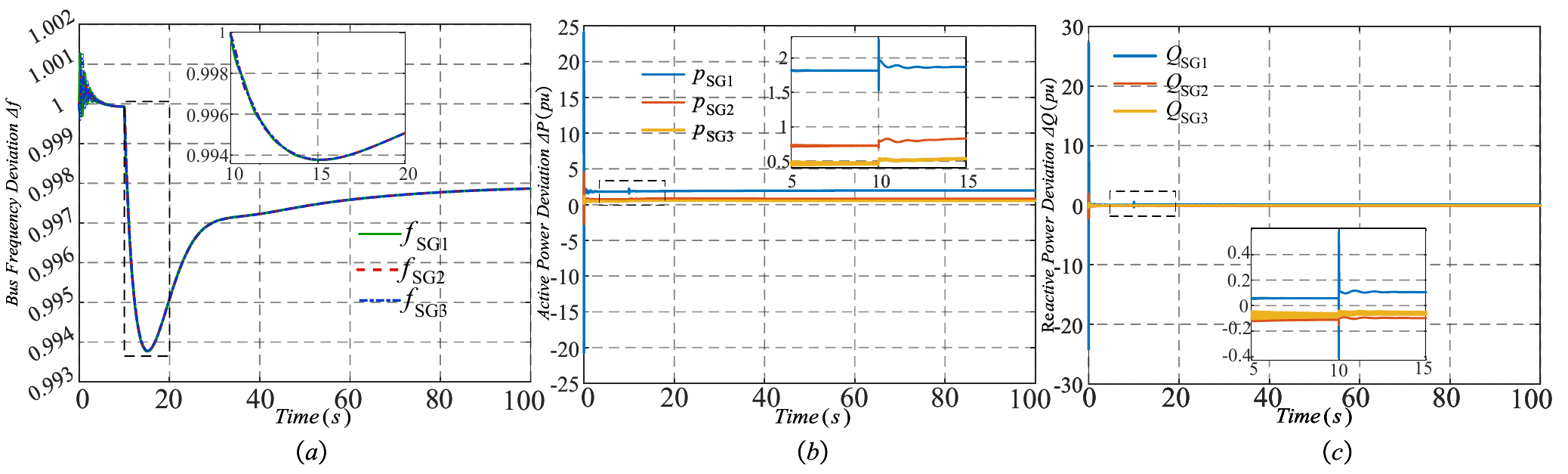}
\caption{Experiment I results: (a) Frequency variations of the three SG units; (b) Active-power deviations of the three SG units; (c) Reactive-power deviations of the three SG units.}
\label{3}
\end{figure}

Overall, the results substantiate the effectiveness of the proposed approach. In Experiment I (without IBRs), Fig.~\ref{3}(a) shows that the three frequency traces nearly coincide after the disturbance, indicating strong system-wide frequency coherence. As shown in Fig.~\ref{3}(b), the active-power response \(\varDelta P\) of the SG fleet is relatively slow with small amplitude; short-term support relies primarily on rotational inertia and primary frequency control, leading to a bandwidth-limited response. Moreover, Fig.~\ref{3}(c) shows negligible changes in \(\varDelta P\), implying weak coupling of this frequency event into the reactive-power channel and a characteristic \(f-p\) dominance.

\begin{figure}[htbp]
\centering
\includegraphics[width=0.48\textwidth]{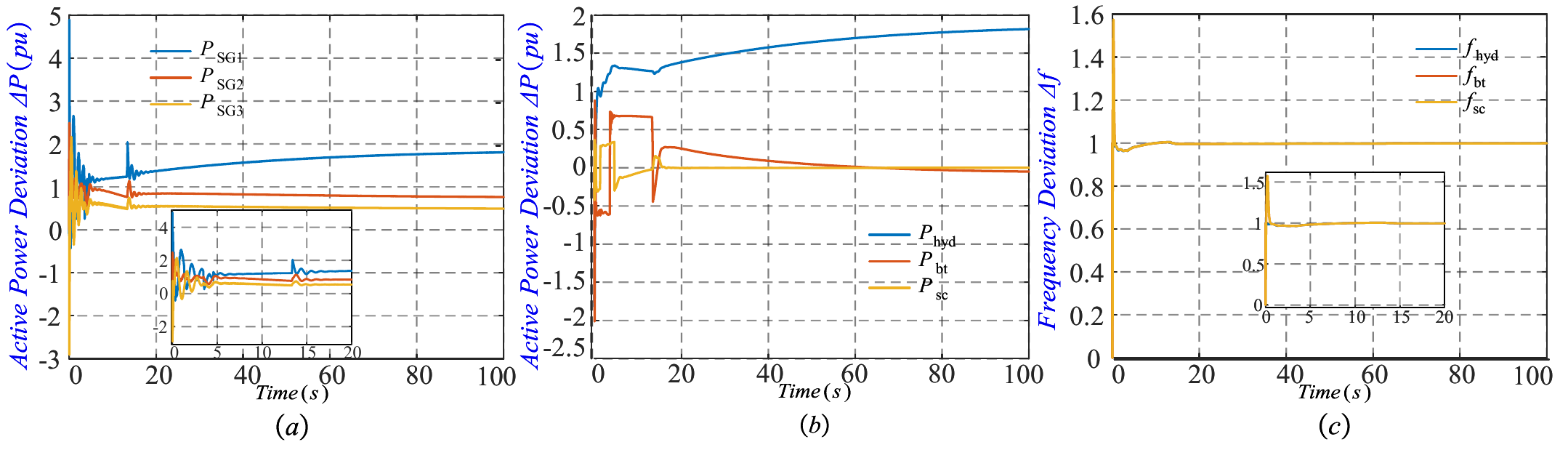}
\caption{Experiment II results: (a) Active-power deviations of the three units; (b) Active-power deviations within DVPP1; (c) Frequency variations within DVPP1.}
\label{4}
\end{figure}

With DVPP1 and the assigned participation factors introduced in Experiment II, Fig.~\ref{4}(a) confirms preserved frequency coherence. Fig.~\ref{4}(b) reveals a clear multi–band relay: the SC delivers prompt high-frequency support, the storage unit provides mid-band regulation, and the hydropower plant supplies the steady-state DC gain; consequently, Fig.~\ref{4}(c) shows a markedly reduced frequency nadir and faster recovery, evidencing effective fast and mid-band support and a suppressed steady-state bias.

\begin{figure}[htbp]
\centering
\includegraphics[width=0.45\textwidth]{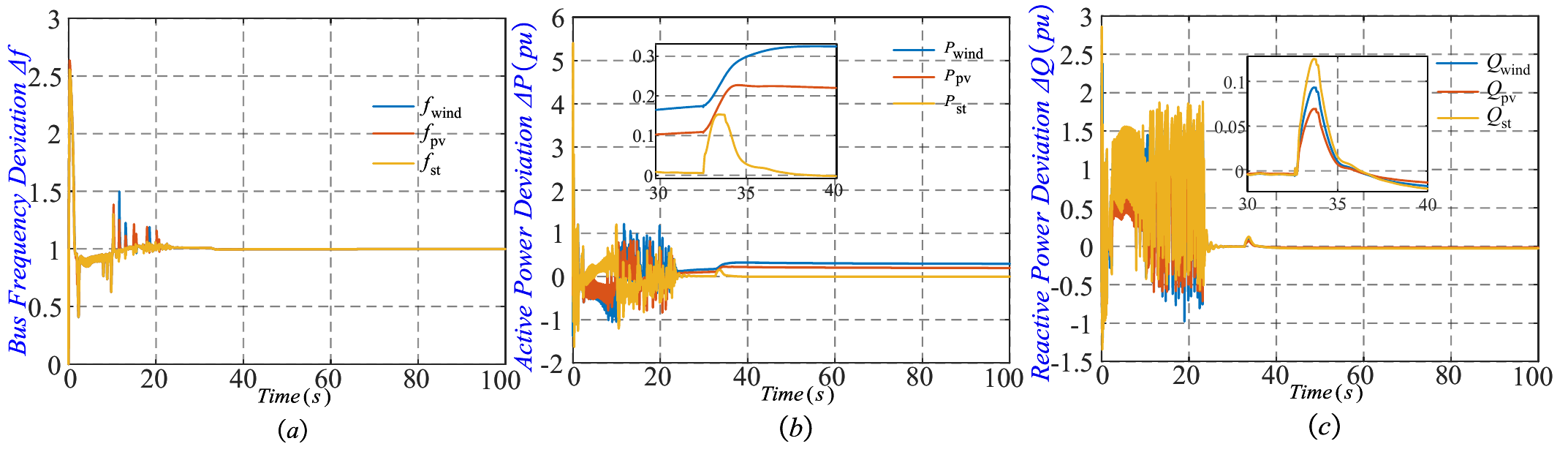}
\caption{Experiment III results: (a) Frequency variations of the three devices within DVPP2; (b) Active-power deviations within DVPP2; (c) Reactive-power deviations within DVPP2.}
\label{5}
\end{figure}

In Experiment III, after integrating the VSG-based DVPP2, Fig.~\ref{5}(a) indicates that the system still experiences a common-frequency event; furthermore, compared with Experiment II, the initial ROCOF is attenuated, reflecting enhanced equivalent inertia and damping from the VSG control. Fig.~\ref{5}(b) demonstrates a clear division of labor on the active-power channel: the vehicle-based storage responds rapidly to high-frequency components, while PV and DFIG assume short-to-intermediate regulation and transition smoothly with storage. The aggregated output closely tracks the target allocation. Meanwhile, Fig.~\ref{5}(c) shows that reactive-power adjustments remain within controllable bounds without noticeable coupling amplification or limit violations, indicating stable voltage support attributable to the GFM voltage-source characteristics. Taken together, the three experiments show that the proposed DVPP achieves coordinated multi–timescale sharing and aggregate matching; with VSG integration, inertia and damping are enhanced, and frequency coherence and operational feasibility are preserved even with renewable participation.

\section{Conclusion}
To ensure that a DVPP exhibits reliable dynamic behavior to the grid and delivers sufficient ancillary services, this paper develops a multi–timescale coordination framework based on grid-forming control. The core idea is threefold. First, a VSG control method is developed to endow the DVPP with quantifiable effective inertia and damping. Second, a dynamic participation–factor decomposition method is constructed along the \(\omega -P\) and \(v-Q\) control loops. Third, a low-/band-/high-pass partitioned coordination strategy is proposed according to device heterogeneity. This enables an orderly task allocation across slow, medium, and fast timescales. Simulation results demonstrate that the proposed method preserves frequency coherency while markedly reducing frequency nadir and initial ROCOF. It can enhance effective inertia and damping, ensure stable reactive power and voltage support, and keep the aggregated output closely aligned with the target allocation. These prove the effectiveness and the implementability of the proposed strategy. Future works include investigating parallel operation of multiple DVPPs, developing adaptive tuning under broader uncertainty, and advancing hardware-in-the-loop validation and practical deployment.


\begin{thebibliography}{00}
\bibitem{b1} V. Häberle, A. Tayyebi, X. He, E. Prieto-Araujo and F. Dörfler, ``Grid-Forming and Spatially Distributed Control Design of Dynamic Virtual Power Plants," \textit{IEEE Transactions on Smart Grid}, vol. 15, no. 2, pp. 1761-1777, March. 2024.
\bibitem{b4} X. He, J. Duarte, V. Häberle, et al., ``Grid-Forming Control of Modular Dynamic Virtual Power Plants,” \textit{arXiv preprint arXiv:2410.14912}, 2024.
\bibitem{b2} V. Häberle, M. W. Fisher, E. Prieto-Araujo and F. Dörfler, ``Control Design of Dynamic Virtual Power Plants: An Adaptive Divide-and-Conquer Approach," \textit{IEEE Transactions on Power Systems}, vol. 37, no. 5, pp. 4040-4053, Sept. 2022.
\bibitem{b3} J. Björk, K. H. Johansson and F. Dörfler, ``Dynamic Virtual Power Plant Design for Fast Frequency Reserves: Coordinating Hydro and Wind," \textit{IEEE Transactions on Control of Network Systems}, vol. 10, no. 3, pp. 1266-1278, Sept. 2023.
\bibitem{b5} W. Zhong, J. Chen, M. Liu, M. A. A. Murad, and F. Milano, ``Coordinated Control of Virtual Power Plants to Improve Power System Short-Term Dynamics," \textit{Energies}, 2021.
\bibitem{b6} A. Tayyebi, D. Groß, A. Anta, F. Kupzog and F. Dörfler, ``Frequency Stability of Synchronous Machines and Grid-Forming Power Converters," \textit{IEEE Journal of Emerging and Selected Topics in Power Electronics}, vol. 8, no. 2, pp. 1004-1018, June. 2020.
\bibitem{b7} G. Cui, Z. Chu and F. Teng, ``Control-Mode as a Grid Service in Software-Defined Power Grids: GFL vs GFM," \textit{IEEE Transactions on Power Systems}, vol. 40, no. 1, pp. 314-326, Jan. 2025.
\bibitem{b8} S. Chen, Y. Wang, Z. Tian, X. Xiao, X. Xie, and O. Gomis-Bellmunt, ``Understanding a type of forced oscillation in grid-forming and grid-following inverter connected systems," \textit{IEEE Transactions on Power Electronics}, vol. 40, no. 8, pp. 11628–11640, Aug. 2025.
\bibitem{b9} Y. Tong, Q. Wang, and A. Tang, ``A novel inverter control strategy with power decoupling for microgrid operations in grid-connected and islanded modes," \textit{arXiv preprint arXiv:2505.06664}, 2025.
\bibitem{b10} Y. Wang, C. Xu, D. Xie, C. Gu, P. Zhao, J. Gong, M. Pan, and X. Wang, ``A novel scheduling strategy for virtual power plant based on power market dynamic triggers," \textit{Applied Energy}, vol. 350, Art. no. 121758, Nov. 2023.
\bibitem{b11} H. Golpîra and B. Marinescu, ``Enhanced Frequency Regulation Scheme: An Online Paradigm for Dynamic Virtual Power Plant Integration," \textit {IEEE Transactions on Power Systems}, vol. 39, no. 6, pp. 7227-7239, Nov. 2024.
\end{thebibliography}
\end{document}